\documentclass[12pt]{article}
\usepackage{graphicx}% Include figure files
\usepackage{dcolumn}% Align table columns on decimal point
\usepackage{bm}% bold math
\begin{document}
\begin{center}
{\bf {\large Improved Quantum Cost for $n$-bit Toffoli Gates}}

\vspace{0.5cm}
Dmitri Maslov and Gerhard W.\ Dueck \\
Faculty of Computer Science\\
University of New Brunswick\\
Fredericton, NB, E3B 5A3, Canada\\
dmitri.maslov@unb.ca, gdueck@unb.ca 
\end{center}

\begin{abstract}
We present an $n$-bit Toffoli gate quantum circuit based on the realization 
proposed by Barenco {\em et al.}, where some of the Toffoli gates in their
construction are replaced with Peres gates. This results in a
significant cost reduction. Our main contribution is a quantum circuit which simulates
the $(m+1)$-bit Toffoli gate with $32m-96$ elementary quantum gates
and one garbage bit which is passed unchanged. This paper is a corrected and 
expanded version of \cite{ar:md}. 
\end{abstract}

\section{Toffoli gate cost}
Toffoli gates are important building blocks in reversible and
quantum circuits.  In fact, a large number of reversible logic
synthesis methods use $n$-bit Toffoli gates (for example,
\cite{ws:mmd, ws:mdmICCAD}).  The $n$-bit Toffoli gates are mentioned
in classical quantum circuits books (such as \cite{bk:nc}) as well as
in journal publications \cite{ar:bbcd}.  The main reason for
popularity of $n$-bit Toffoli gates over the other gates is their
completeness and relative simplicity in using them.  The high
demand in $n$-bit Toffoli gate makes it important to have a low cost
quantum circuit for this gate.

Barenco {\em et al.} \cite{ar:bbcd} considered all one qubit gates and all 
controlled-V gates \cite{bk:nc} to be elementary. The authors proposed a circuit 
for the $(m+1)$-bit Toffoli gate with a cost of $48m-116$ basic operations plus one 
garbage bit. Here are their two main results (slightly reformulated to 
fit the notations of this paper).

{\em Lemma 7.2.} If number of bits in the circuit $n\geq 5$ and 
$m\in \{3,...,[n/2]\}$, then an $(m+1)$-bit Toffoli gate can be simulated 
by a network consisting of $4(m-2)$ Toffoli gates.

{\em Corollary 7.4.} On an $n$-bit network (where $n\geq 7$), an 
$(n-1)$-bit Toffoli gate can be simulated by $8(n-5)$ Toffoli gates,
as well as by $48n-212(=48(n-2)-116)$ basic operations.

\begin{figure}[h]
\centering
\includegraphics[height=22mm]{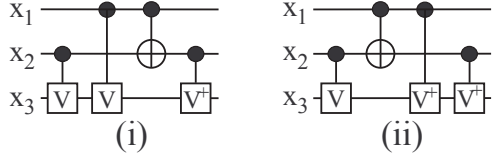}
\caption{Structure of the Peres gate and its inverse}
\label{peres}
\end{figure}

Before we can describe our improved design, we have to introduce the Peres gate.
The Peres gate $P(x_1,x_2,x_3)$ \cite{ar:peres, ws:plpk} is equivalent to the
transformation produced by a Toffoli gate $TOF(x_1,x_2,x_3)$ followed
by a CNOT gate $TOF(x_1,x_2)$.  A four elementary quantum
transformations realization of Peres gate is illustrated in Figure
\ref{peres}(i), where ${\bf V}=\frac{i+1}{2}\left(1\;\;-i\atop -i\;\;1\right)$
and ${\bf V}^+$ is its inverse.  Denote the four gates used in the proposed
construction as $A, B, C$ and $D$.  Trivial analysis shows that the
inverse Peres gate can be achieved by a circuit
$D^{-1}C^{-1}B^{-1}A^{-1}$ (Figure \ref{peres}(ii)), consisting of the
inverses of the gates used for construction of Peres gate.  From the
point of view of Toffoli-CNOT realization, the inverse Peres gate will
act as a CNOT $TOF(x_1,x_2)$ followed by the Toffoli gate
$TOF(x_1,x_2,x_3)$.

\begin{figure}[htb]
\centering
\includegraphics[height=44mm]{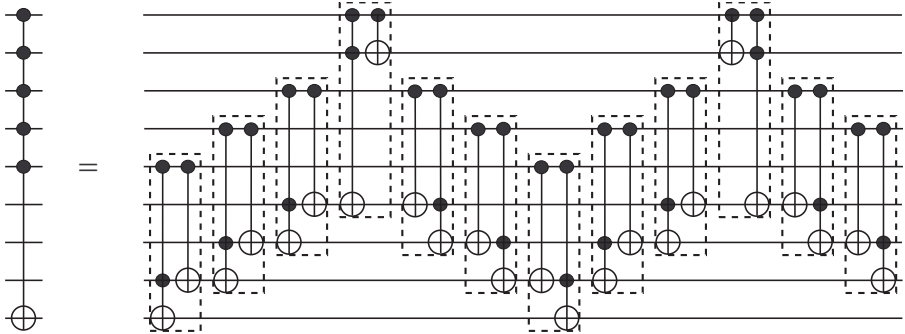}
\caption{Circuit for $(m+1)$-bit Toffoli (illustrated for $m=5$)}
\label{lemma72}
\end{figure}

We suggest that in construction of Lemma 7.2 in \cite{ar:bbcd} the
Peres gate or its inverse are used everywhere instead of the more
expensive Toffoli gate.  This is illustrated in Figure
\ref{lemma72}, where each of the pairs Toffoli-CNOT and CNOT-Toffoli
is a Peres gate or its inverse.

To prove that such circuit realizes an $(m+1)$-bit Toffoli one can
inspect it or simply notice that a pair of identical CNOTs can be
moved together using the moving rule from \cite{ws:mdmIWLS} and thus,
be canceled out.  Therefore, this circuit becomes equivalent to the one
proposed by Barenco {\em et al.} \cite{ar:bbcd} which was shown to
simulate a Toffoli gate.  With the Peres gates the network for
$(m+1)$-bit Toffoli gate will have a cost of $4*(4m-10)+4*2=16m-32$
elementary quantum operations plus $(m-2)$ garbage bits.  Using our
construction in Corollary 7.4 of \cite{ar:bbcd} one can
achieve a cost of $32m-96$ elementary operations plus one garbage bit
for $(m+1)$-bit Toffoli gate construction for $m\geq 5$, which is
better than the calculated in \cite{ar:bbcd} $48m-116$ elementary
operations plus one garbage bit.

Further, we propose to use the Peres gate in all similar
constructions (for example, the circuit on page 184 of \cite{bk:nc})
for a better quantum cost analysis.

\section{Usefulness of the results}

Barenco {\em et al.} \cite{ar:bbcd} among other results provide the following Lemma (slightly reformulated to fit the %%@
context of the presented paper).

{\em Lemma 7.1.} For any $m\geq 2$, $(m+1)$-bit Toffoli gate can be simulated with $2^{m+1}-3$ controlled-V quantum %%@
operations. 

Since for small numbers $m \;\;2^{m+1}-3$ can be less than $16m-32$ or $32m-96$ achieved here (and less than %%@
$4(m-2)T$, where T is the cost of the original Toffoli gate), we would like to illustrate where the new results %%@
improve the known ones. Table \ref{tab:res} summarizes usefulness and applicability of the design that uses the Peres %%@
gate; symbol $``*"$ indicates when the result was achieved using the proposed construction. It is interesting to %%@
notice that the proposed construction which uses Peres gates in Lemma 7.2 of \cite{ar:bbcd} first worked to decrease %%@
quantum cost of size 6 Toffoli gate, even though the construction is valid for smaller Toffoli gates. The first time %%@
formula $32m-96$ updated the known cost estimates is for the size 10 Toffoli gate quantum cost. 

\begin{table}[tb]
    \begin{center}
    \begin{tabular}{|c|c|c|} \hline
{\bf Toffoli gate size $(m+1)$} & {\bf Garbage}	& {\bf Cost}   \\ \hline
1 & 0 & 1 \\
2 & 0 & 1 \\
3 & 0 & 5 \\
4 & 0 & 13 \\
5 & 0 & 29 \\
6 & 0 & 61  \\
6 & 1 & 52  \\
6 & 3 & 48*  \\
7 & 0 & 125  \\
7 & 1 & 84  \\
7 & 4 & 64*  \\
8 & 0 & 253  \\
8 & 1 & 116  \\ 
8 & 5 & 80*  \\
9 & 0 & 509  \\
9 & 1 & 154*  \\
9 & 6 & 96*  \\
10 & 0 & 1021  \\
10 & 1 & 192*  \\
10 & 7 & 112*  \\
$(m+1)>10$ & 0 & $2^{m+1}-3$  \\
$(m+1)>10$ & 1 & $32m-96*$  \\
$(m+1)>10$ & $m-2$ & $16m-32*$  \\ \hline
\end{tabular}
\end{center}
\caption{Quantum costs of the Toffoli gates with $m$ controls}
\label{tab:res}
\end{table}   

\bibliographystyle{plain}
\bibliography{rl}
%\newpage
%\noindent {\bf Authors' affiliations:}
%
%\noindent Dmitri Maslov and Gerhard W. Dueck (Faculty of Computer Science, University 
%of New Brunswick, P.O. Box 4400, Fredericton, NB, E3B 5A3, Canada)
%\newline
%\newline
%\noindent E-mail: cetrau@mail.ru
%
%\newpage
%
%\noindent {\bf Figure captions:}
% 
%\noindent Fig. 1 Structure of the Peres gate and its inverse
%
%\noindent Fig. 2 Circuit for $(m+1)$-bit Toffoli (illustrated for $m=5$)
%
%\newpage
%
%\noindent {\bf Figure 1}
%
%\begin{figure}[h]
%\centering
%\includegraphics[height=20mm]{peres.eps}
%%\caption{Structure of the Peres gate and its inverse}
%\label{peres}
%\end{figure}
%
%\newpage
%
%\noindent {\bf Figure 2}
%
%\begin{figure}[htb]
%\centering
%\includegraphics[height=40mm]{lemma72.eps}
%%\caption{Circuit for $(m+1)$-bit Toffoli (illustrated for $m=5$)}
%\label{lemma72}
%\end{figure}
%
\end{document}